\documentclass[10pt, conference]{IEEEtran}
\IEEEoverridecommandlockouts
\usepackage{amsmath,amssymb,amsfonts}
\usepackage{algorithmic}
\usepackage{graphicx}
\usepackage{tabularx}
\usepackage{textcomp}
\usepackage{xcolor}
\usepackage{siunitx}
\usepackage{booktabs}
\usepackage{url}
\usepackage{enumitem}
\usepackage{diagbox}
\usepackage{multicol}
\usepackage{multirow}
\usepackage{threeparttable}
\usepackage[numbers]{natbib}
\usepackage[listings]{tcolorbox}
\usepackage{multirow}
\usepackage{array} 

\usepackage{times}
\usepackage{graphicx}
\usepackage{epsf}
\usepackage{verbatim}
\usepackage{url}
\usepackage{color}
\usepackage{alltt}

\newcommand{\Caption}{\caption}

\newcommand{\CodeIn}[1]{{\small\texttt{#1}}}

\newcommand{\Comment}[1]{}

\newenvironment{CodeOut}{\begin{scriptsize}}{\end{scriptsize}}

\newcommand{\SmallSpace}{\vspace*{-1.4ex}}

\usepackage{braket}
\usepackage{pifont}% http://ctan.org/pkg/pifont

\def\BibTeX{{\rm B\kern-.05em{\sc i\kern-.025em b}\kern-.08em
    T\kern-.1667em\lower.7ex\hbox{E}\kern-.125emX}}
\begin{document}

\title{An Empirical Study of Bugs in Quantum Machine Learning Frameworks\\
}

%\author{Anonymous*}

\author{
\IEEEauthorblockN{Pengzhan Zhao,
Xiongfei Wu,
Junjie Luo,
Zhuo Li,
Jianjun Zhao\IEEEauthorrefmark{1}
\thanks{\IEEEauthorrefmark{1}zhao@ait.kyushu-u.ac.jp} 
}
\IEEEauthorblockA{
Graduate School and Faculty of Information Science and Electrical Engineering\\
Kyushu University, Japan\\
%Email: {zhao@ait.kyushu-u.ac.jp}
}
}

% \and
% \and
% \and
% \and
% \and
% \textit{name of the organization (of Aff.)}\\

\maketitle

% \begin{abstract}
% Recent advances in quantum computing have inspired the innovation of quantum machine learning (QML) frameworks, such as IBM Qiskit Machine Learning and TensorFlow Quantum.  Given the importance of quantum machine learning, we seek to answer the critical question to ease the adoption and development of QML frameworks: \emph{What challenges do users face when using QML frameworks, and what are the common challenges for developers when developing QML frameworks?} In this paper, we present the first empirical study to identify the challenges in both the usage and development of QML frameworks. We choose \textcolor{red}{nine} most popular QML frameworks in the market and manually inspect \textcolor{red}{1591} bug reports collected from their corresponding GitHub repositories. We characterize QML bugs in a total of \textcolor{red}{yy} impacts to obtain an initial understanding of defects in quantum machine learning frameworks. Furthermore, we distill a taxonomy of challenges of using QML frameworks consisting of \textcolor{red}{zz} categories and \textcolor{red}{cc} types of common topics about developing QML frameworks. Finally, we propose \textcolor{red}{??} implications for researchers and developers to improve the development practices further and to build more robust QML frameworks.
% \end{abstract}

\begin{abstract}
Quantum computing has emerged as a promising domain for the machine learning (ML) area, offering significant computational advantages over classical counterparts. With the growing interest in quantum machine learning (QML), ensuring the correctness and robustness of software platforms to develop such QML programs is critical. A necessary step for ensuring the reliability of such platforms is to understand the bugs they typically suffer from. To address this need, this paper presents the first comprehensive study of bugs in QML frameworks. We inspect 391 real-world bugs collected from 22 open-source repositories of nine popular QML frameworks. We find that 1) 28\% of the bugs are quantum-specific, such as erroneous unitary matrix implementation, calling for dedicated approaches to find and prevent them; 2) We manually distilled a taxonomy of five symptoms and nine root cause of bugs in QML platforms; 3) We summarized four critical challenges for QML framework developers. The study results provide researchers with insights into how to ensure QML framework quality and present several actionable suggestions for QML framework developers to improve their code quality.
\end{abstract}

\begin{IEEEkeywords}
quantum machine learning, quantum software testing, quantum program debugging, empirical study
\end{IEEEkeywords}

\section{Introduction}\label{sec:intro}
% Machine learning (ML) techniques have been widely applied to many cutting-edge areas, e.g., machine translation~\cite{wang2020tencent}, cancer diagnosis~\cite{fakoor2013using}, and autonomous cars~\cite{BADUE2021113816}. 
Quantum computing has been making immense progress due to increasing computer power and algorithmic advances~\cite{cerezo2021,farhi2022algorithm}. Tremendous efforts from industry and academia have greatly stimulated the evolution of this area. Although fault-tolerant quantum computers will likely not be available shortly, recent research on quantum machine learning has revealed the potential of quantum computers to outperform classical computers on machine learning tasks~\cite{Biamonte_2017}. 

With the rapidly growing complexity of quantum programs, it is decisive to alleviate the efforts in programming such quantum programs. \emph{Quantum computing framework} provides the essential interface (i.e., quantum programming language), compiler, and execution environment for quantum programmers to run quantum programs on a quantum computer or a simulator. Several quantum programming frameworks are available for quantum programmers, such as Qiskit~\cite{aleksandrowicz2019qiskit} by IBM, Q\#~\cite{svore2018q} by Microsoft, and Cirq~\cite{cirq2018google} by Google, allowing researchers and developers to implement and experiment with various quantum programs quickly. Furthermore, stimulated by the prosperity of the classical ML community and the potential of QML techniques, a number of QML frameworks are proposed, such as Torch Quantum~\cite{hanruiwang2022quantumnas} and PennyLane~\cite{bergholm2018pennylane}. Given the importance of this rapidly involving field, ensuring the correctness of the underlying QML frameworks deserves high priority. Various approaches exist to prevent and find bugs in quantum computing frameworks, e.g., for bug characteristics~\cite{paltenghi2022bugs, zhao2021bugs4q} or testing~\cite{matteo2022morphq1, wang2022qdiff}. However, little attention has been received to ensure the quality of QML frameworks. One efficient approach to help prevent and detect bugs is understanding and characterizing bugs that exist in the wild~\cite{paltenghi2022bugs}. However, there currently is no detailed study of bugs in QML frameworks.

To bridge this gap, we present the first empirical study to characterize bugs in QML frameworks. We collect and inspect a set of 391 real-world bugs from 22 open-source projects, including highly popular repositories such as Torch Quantum and PennyLane. We aim to answer three fundamental questions that remain unclear: \emph{How many of these bugs are specific to quantum computing, where do these bugs occur in QML platforms, and how do these bugs manifest?} Furthermore, we identify key challenges QML developers face when developing QML platforms. The key findings of this study include the following:

\begin{itemize}
    \item 28\% of the bugs are quantum-specific, and these quantum-related bugs are primarily associated with qubit manipulation, such as erroneous unitary matrix implementations, calling for dedicated approaches to find and prevent them.
    \item While the most prevalent symptom is program crash, quantum-related bugs tend to produce function errors, which are challenging to detect. Moreover, QML frameworks are more prone to environmental factors, such as framework version and device compatibility, which may cause warning errors.
    \item Root cause analysis reveals that algorithmic and logic issues accounted for the largest proportion, with inconsistencies stemming from version lags and updated features as the second most common cause.
    \item Based on the bug patterns, we present four key challenges that QML framework developers should address. For instance, ensuring cross-device execution of QML programs, given the multitude of devices that support them (e.g., GPUs, QVMs, QPUs, and simulators).
\end{itemize}

The results of this study provide insights for both researchers and developers. Researchers may benefit from insights into the bugs in this domain and can propose new approaches to tackle the problems they cause. QML framework developers can learn the bug patterns and be careful when facing the key challenges we propose. To better help other research in this domain, we publicly share our QML framework bug dataset, including annotation intermediate procedure and results.

In summary, this paper contributes the following:
\begin{itemize}[leftmargin=2em]
\setlength{\itemsep}{3pt}
  % \item We distill a set of common bug patterns for quantum programs, which is the foundation of performing static analysis on quantum programs.
  \item The first study of 391 real-world bugs in 22 popular QML frameworks.
  
  \item We manually inspected 391 bugs and summarized five symptoms and nine root causes of bugs to help characterize bugs in QML frameworks.

  \item Insights of challenges developers face when developing QML frameworks.

  \item A publicly available dataset of labeled, real-world bugs to support future research in testing QML frameworks.
\end{itemize}

The rest of the paper is organized as follows. Section~\ref{sec:background} provides some basics of quantum programming and quantum machine learning. 
Section~\ref{sec:methodology} describes our methodology for collecting and filtering bugs. Section~\ref{sec:empirical} presents the empirical results of our study. Section~\ref{sec:threats} reviews our threats of validity.
Section~\ref{sec:related} discusses related work, and Section~\ref{sec:conclusion} finally concludes this paper.

\section{Background}\label{sec:background}
This section briefly introduces some basic concepts of quantum computing
as well as quantum machine learning. 

\subsection{Quantum Computing}

A quantum bit (qubit) is the analog of one classical bit but has many different properties. A classical bit, like a coin, has only two states, 0 and 1, while a qubit can be in a continuum of states between $\Ket{0}$ and $\Ket{1}$ in which the $\Ket{}$ notation is called Dirac notation. We can represent a qubit mathematically as $\Ket{\psi} = \alpha\Ket{0} + \beta\Ket{1}$ where $|\alpha|^2 + |\beta|^2 = 1$ and the numbers $\alpha$ and $\beta$ are complex numbers. The states $\Ket{0}$ and $\Ket{1}$ are computational basis states. Unlike classical bits, we cannot examine a qubit directly to get the values of $\alpha$ and $\beta$. Instead, we measure a qubit to obtain either 0 with probability $|\alpha|^2$ or 1 with probability $|\beta|^2$.

Quantum gates are used for quantum computation, which means manipulating quantum information. Some basic quantum gates are listed as follows:
\begin{itemize}[leftmargin=2em]
\setlength{\itemsep}{3pt}
  \item Quantum NOT gate takes the state $\Ket{\psi} = \alpha\Ket{0} + \beta\Ket{1}$ into the state $\Ket{\psi} = \alpha\Ket{1} + \beta\Ket{0}$. We can use a matrix to represent this operation:
  \begin{gather*}X = \begin{bmatrix} 0 & 1\\1 & 0\end{bmatrix}\end{gather*}
  \item The Z gate can be expressed as 
  \begin{gather*}Z = \begin{bmatrix} 1 & 0 \\ 0 & -1\end{bmatrix}\end{gather*} 
  From the matrix, we know the Z gate leaves the $\Ket{0}$ unchanged and changes the sign of $\Ket{1}$.
  \item The Hadamard gate turns the $\Ket{0}$ into $(\Ket{0} + \Ket{1})/\sqrt{2}$ and turns the $\Ket{1}$ into $(\Ket{0} - \Ket{1})/\sqrt{2}$. The matrix form of the Hadamard gate is 
  \begin{gather*}H = \frac{1}{\sqrt{2}}\begin{bmatrix} 1 & 1\\ 1 & -1 \end{bmatrix}\end{gather*}
\end{itemize}

All the matrices are unitary ones. Besides these single-qubit gates, there are multiple qubit gates, such as the Controlled-NOT gate (CNOT gate). This gate has two input qubits, the control and target qubits. If the control qubit is 0, then the target qubit remains unchanged. If the control qubit is 1, then the target qubit is flipped. We can express the behavior of the CNOT gate as $\Ket{A, B} \rightarrow \Ket{A, B \oplus A}$. 
\iffalse
The matrix representation is 
\begin{gather*}U_{CN} = \begin{bmatrix}  1 & 0 & 0 & 0 \\ 0 & 1 & 0 & 0 \\ 0 & 0 & 0 & 1 \\ 0 & 0 & 1 & 0 \end{bmatrix}\end{gather*}
\fi

Quantum circuits are models of all kinds of quantum processes. We can build quantum circuits with quantum gates and use wires to connect the components in quantum circuits. These wires can represent the passage of time or a physical particle moving from one position to another. Another essential operation in quantum circuits is measurement. Measurement operation observes a single qubit and obtains a classic bit with a certain probability. 
Nielsen's book\cite{nielsen2002quantum} has a more detailed explanation of quantum computation.

\subsection{Quantum Machine Learning}\label{sec:qml}

Machine learning is based on statistics and extracts the appropriate patterns by analyzing existing data so that the patterns can be used to make predictions about new data. Machine learning is divided into three main categories, supervised learning, unsupervised learning, and reinforcement learning. Among them, supervised learning and unsupervised learning are mainly distinguished by the input dataset. The supervised learning dataset contains data and corresponding labels and is concerned with the mapping rules from data to labels. On the other hand, the unsupervised learning dataset does not have labels and performs the task of classification by extracting features common to the data. Reinforcement learning focuses on the best decision for a problem by enabling an agent to learn strategies as it interacts with its environment to maximize the reward or achieve the appropriate goal. Machine learning has been widely adopted in various fields, including healthcare, education, manufacturing, etc~\cite{jordan2015machine}.

Quantum machine learning is a rapidly growing field combining quantum algorithms and machine learning principles. Quantum algorithms have been shown to outperform the best classical algorithms for specific problems by exploiting quantum properties such as superposition states, coherence, and entanglement effects, which are known as quantum speedup~\cite{Biamonte_2017}. It has led to the exploration of applying quantum algorithms to machine learning with the expectation that algorithms superior to classical machine learning can be discovered. Although there is still a lack of sufficiently powerful quantum computer hardware to support practical applications of quantum machine learning algorithms, some quantum machine learning ideas have been proposed and validated, such as the quantum algorithm for linear systems of equations (also called the HHL algorithm)~\cite{Harrow_2009}, the Variational Quantum Eigensolver (VQE)~\cite{Peruzzo_2014}, and the Quantum Convolutional Neural Network (QCNN)~\cite{Cong_2019}, etc. \

Overall, the field of quantum machine learning is still in its early stages. However, it holds great potential for pushing the boundaries of traditional machine learning research and solving complex problems that are difficult or even impossible to tackle with classical methods.

\section{Methodology}\label{sec:methodology}
In this section, we describe how we conducted our study by formulating research questions,  how we selected the study object and data source, how we constructed the dataset, and how we performed the study.

\subsection{Research Questions}
This study is driven by the following three research questions:
\begin{itemize}[leftmargin=2em]
\setlength{\itemsep}{3pt}
  % \item We distill a set of common bug patterns for quantum programs, which is the foundation of performing static analysis on quantum programs.
  \item \textit{RQ1:Where the bugs occur and how many of the bugs in QML platforms are specific to quantum computing?}
  This question represents our first purpose, which is to understand the relationship between bugs in the QML frameworks and traditional ML frameworks. 
  
  \item \textit{RQ2:What are the symptoms and causes of bugs in QML platforms?}
  This question is the core and foundation of this study, and understanding the symptoms and causes of the platform can help better detect and fix bugs. In addition, it also helps to answer RQ3.
  %\item \textit{RQ3:What challenges do users face when using QML frameworks?}
  %Due to the multi-platform and multi-device implementation of QML, users are faced with more than simply using the QML platforms.
  %\item \textit{RQ3:What are the common challenges for developers when developing QML frameworks?}
  \item \textit{RQ3:What are the areas of conflict that developers need to handle when developing QML frameworks?}
    Developers can also face many challenges when refining the QML platform. Especially for QML framework developers, not only do they need to implement ML functionality, but they also need to implement quantum operations.
\end{itemize}
\subsection{Dataset}
% 22 repositories
The main objective of this paper is to study bugs in quantum machine learning frameworks. Following previous works~\cite{paltenghi2022bugs, zhao2021bugs4q}, we hence study not only a single or a few projects but 22 open-source projects in total. Furthermore, we decided to focus on open-source GitHub projects, as these projects allow us to apply the same process procedure to each project. To select projects, we first utilize the list of quantum computing frameworks used in~\cite{paltenghi2022bugs} and check whether these projects contain a corresponding quantum machine learning framework. If so, the corresponding quantum machine learning framework is considered to be added to the initial list. Then we extend the initial list by inspecting the search results of ``quantum machine learning'' and ``quantum neural network'' on Google Scholar and GitHub, which contain a corresponding GitHub repository. As a result, 22 popular quantum machine learning frameworks are selected. The selected projects cover all the quantum machine learning techniques introduced in Section~\ref{sec:qml} (e.g., QCNN). Due to limited space, detailed information about the selected projects can be found at our anonymous website\footnote{https://anonymous.4open.science/r/QML-platform-bugs-0E90/}.

%\zpz{The collection results are shown in Table~\ref{table:1}.} 
After selecting QML frameworks, we crawled all the issue reports from the 22 repositories above, consisting of 1,591 issue reports. Then we manually inspected these reports to filter out the candidate bug reports to prevent introducing noise not about bugs in QML frameworks. For instance, sometimes users may think of a specific behavior as a bug in QML frameworks, which may be an API misuse; some posts were just about typos in the tutorial documents; and some posts indeed discussed bugs but did not yet obtain a correct answer, which could not be used to help characterize bugs in QML frameworks. During the inspection process, we kept track of all the materials (e.g., external links, commit messages, etc.) in the issue report to obtain a complete picture. Overall, we identify 391 bugs to study and answer our RQs.

\subsection{Understanding and Labeling Properties of Bugs}
Answering our research questions requires a solid understanding of the identified bugs. This section describes how we inspect and label the bugs with various properties.

\subsubsection{Traditional vs. Quantum-Specific} To address RQ1, we need to classify each bug as either traditional or quantum-specific. Following previous work~\cite{paltenghi2022bugs}, we consider a bug \textit{quantum-specific} if the mistake involved is related to quantum-specific concepts. Specifically, understanding these bugs requires knowledge of the quantum programming domain. Otherwise, we consider all other bugs as traditional, which means that these bugs may occur in traditional ML frameworks and do not need quantum-specific knowledge to be fixed.

\subsubsection{Identifying Symptoms and Root Causes} Addressing RQ2 and RQ3 requires us to label all bugs concerning two dimensions: symptom and their corresponding root cause. Following previous ML bug studies~\cite{yang2022ist}, two authors started with the classification schema used for labeling from the existing general ML bug studies~\cite{thung2012bugs, yang2022ist}. They adapted it by appending new ones and excluding unrelated ones. To carefully annotate bugs, they read all the bug reports separately, including the title, description, comments, and the corresponding code changes.

In particular, the symptom of a bug is determined if the developer explicitly reported the symptom in the context of the reports. The root cause of a bug was inferred from the discussion as well as the corresponding code changes. The root cause of a bug will be labeled as ``Unclear'' if there are no explicit answers or code patches of a bug to avoid introducing bias.

In summary, we carefully inspected 391 bug reports and distilled a total of five symptoms and nine root causes. A third author was involved in resolving disagreements during the labeling procedure. The Cohen's Kappa coefficient for symptoms and root causes was 0.73 and 0.7, which implies substantial agreement. The complete inspection and labeling procedure requires about 500 person-hours.

\section{Empirical Results}\label{sec:empirical}
This section details the results of our study as well as the answers to our three research questions.
\subsection{RQ1.Locations and Quantum-specific bugs}

A quantum platform bug occurs in the source code, which belongs to the various components of the QML platform.
The components and names of the QML platform vary due to the different frameworks. We group the components into the following categories based on statistical bugs and a comparison of the various platforms:
\begin{enumerate}[itemindent=1.5em]
    \item \textbf{Quantum basic.} Quantum basic is mainly concerned with the most basic functions for qubit states. In particular, most of these components for building basic operations on qubits are often in the basic library of one framework such as Cirq\footnote{https://github.com/quantumlib/Cirq} or Qiskit-Terra\footnote{https://github.com/Qiskit/qiskit-terra} rather than in QML libraries. Therefore, most bugs found in the quantum basic-related files exist in the \CodeIn{pennylane/measurements} and  \CodeIn{pennylane/pauli} directory, etc. 
   
    \item \textbf{Kernel.} This component implements the quantum kernel matrices that can be used to implement classical machine learning algorithms. All bugs of this location type are contained in the \CodeIn{quantum\_kernel.py} file and the \CodeIn{pennylane/kernels} directory.

    \item \textbf{Drawer.} This is a graphical feature to visualize quantum circuits. As with \textit{quantum basic}, this is a feature unique to Pennylane. Some bugs are located in file \CodeIn{pennylane/circuit\_graph.py} or directory \CodeIn{pennylane/drawer}.

    \item \textbf{Template.} This component mainly provides templates that can be used in some quantum nodes, such as quantum state preparation and layers for transforming quantum states. We locate this component in \CodeIn{qiskit\_machine\_learning/neural\_networks} and \CodeIn{pennylane/templates} directory.

    \item \textbf{Interface.} QML programs rely heavily on other equipment. And many of the interfaces, led by Pennylane, are not only hardware-oriented but also have simulator backends for other platforms.
    Many bugs lead to the programs failing to run on GPUs. An example can be seen in issue \#1705\footnote{https://github.com/PennyLaneAI/pennylane/pull/1705}. This component is located in \CodeIn{pennylane/devices} and \CodeIn{pennylane/interfaces}.

    \item \textbf{Algorithm.} Many QML algorithms (e.g. VQE, VQC, and QAOA) are integrated with various frameworks. This component can be located in \CodeIn{pennylane/fourier} and \CodeIn{qiskit\_machine\_learning/algorithms} etc.

    \item \textbf{Transform.} The \CodeIn{pennylane/transforms} directory provides the functions to convert and decompose the large code base and supply to help achieve the compilation of quantum circuits.

    \item \textbf{Operation.} This component contains the core quantum operations, mainly in the form of unitary operations. It is different from \textit{quantum basic}, which mainly focuses on the composition of quantum elements. We can locate the \textit{ops} in directory \CodeIn{tensorflow\_quantum/core/ops/} and directory \CodeIn{pennylane/ops}.

\end{enumerate}

\begin{figure}[h]
    \centering
    \includegraphics[width=8.4cm]{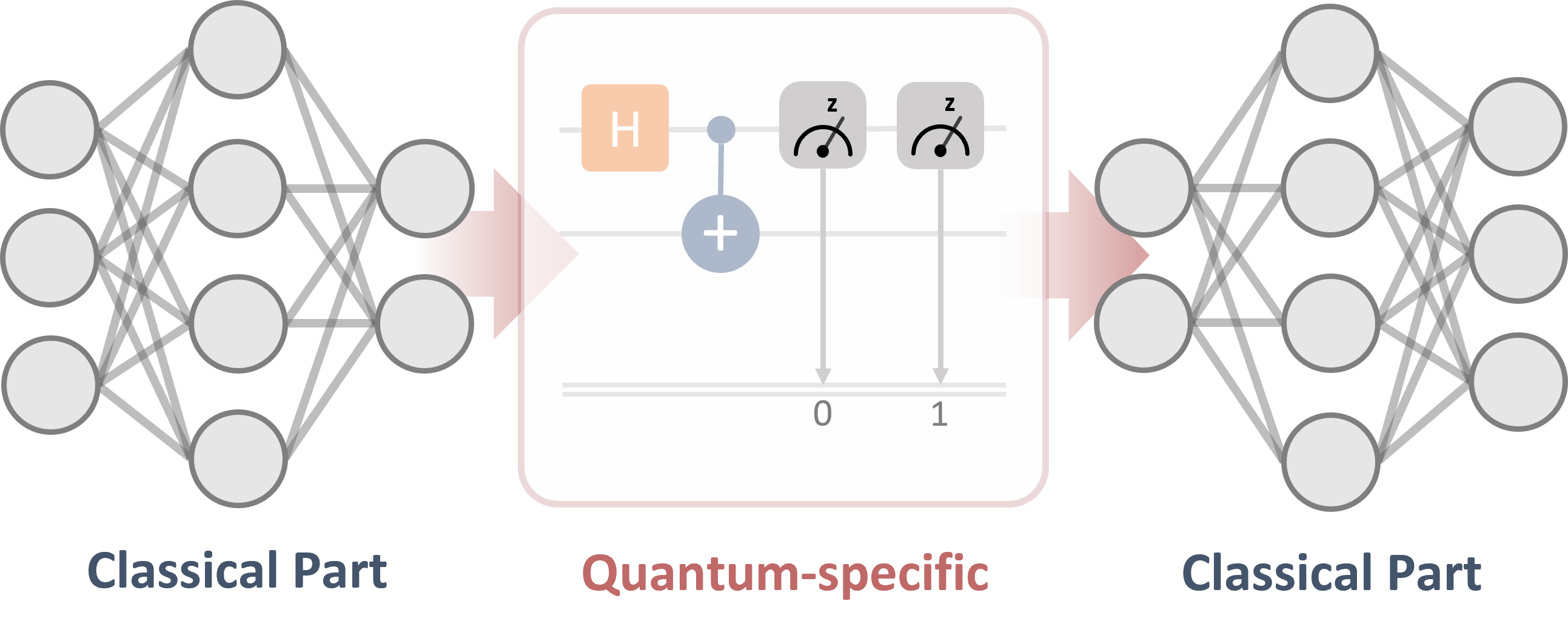}
    \caption{Structure of QML programs}
    \label{fig:qs}
    %\vspace{-2mm}
\end{figure}
Figure~\ref{fig:qs} shows a simple structure of QML programs.
Here is a disclaimer: Since all quantum platforms at this stage are based on classical programming languages to simulate the implementation of qubit state transitions, all bug symptoms and causes fall under the category of classical programming.
If a bug occurs at the quantum part, we categorize it as a quantum-specific
bug. An example can be seen in Pennylane \#3100\footnote{https://github.com/PennyLaneAI/pennylane/issues/3100}, which shows a function error of multiple-qubits gates. And the cause of this bug is incorrect logic being used. Figure~\ref{fig:2} shows a part of the modification.
\begin{figure}[t]
  \begin{CodeOut}
\footnotesize{
  \begin{alltt}
    \textcolor{purple}{-       if op.num_wires != 2:}
    \textcolor{teal}{+       if len(op.wires) == 1:}
                gates.append(op)
                list_op_copy.pop(0)
                continue
  \end{alltt}
}
\end{CodeOut}
\Caption{\label{fig:2} Part of modified code}
\end{figure}
As a result, a quantum-specific bug is reflected in the elements that implement the quantum circuit part, e.g., the encoding of qubits, quantum gates for conversing qubit operations, and the function to implement measurement, etc.
\iffalse
\begin{figure}[h]
    \centering
    \includegraphics[width=6.5cm]{Fig/1.png}
    \caption{Distribution of bugs}
    \label{fig:321}
    % \vspace{-9mm}
\end{figure}
\fi

\begin{figure}[t]
    \centering
    \includegraphics[width=0.46\textwidth]{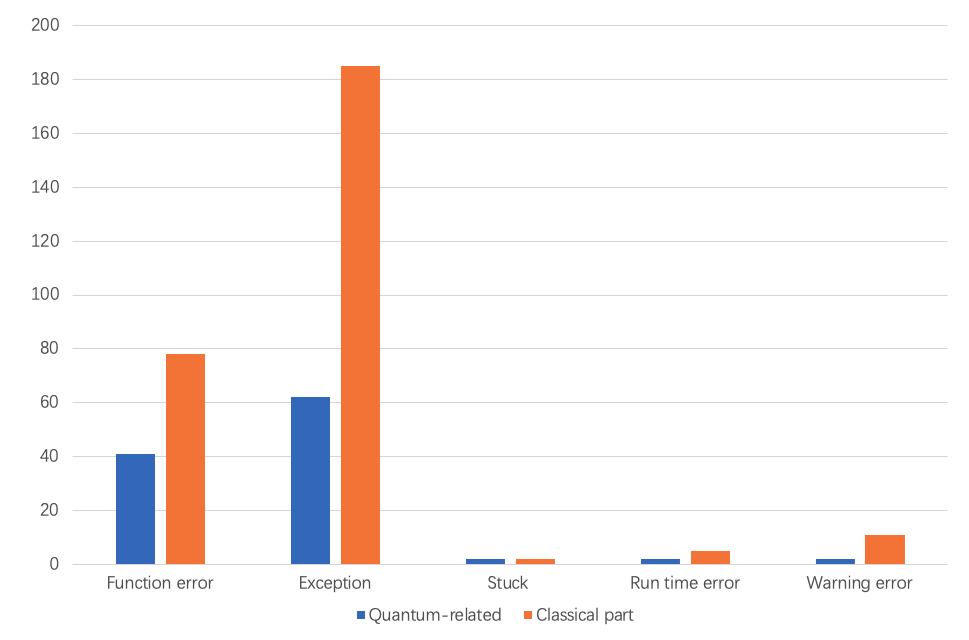}
    \caption{The proportion of quantum-related bugs and traditional bugs in each symptom.}
    \label{fig:type}
    \vspace{-2mm}
\end{figure}

\textbf{Finding 1.}
%Figure~\ref{fig:321} shows the result of our collection.
There are 108 quantum-related bugs, accounting for 28\% of the total bug count of 391, while the remaining 72\% of bugs occur on the classical part. 
Quantum-specific bugs are all reflected in the functionality of qubit manipulation, including state vector misrepresentation of qubits and error implementation of the unitary matrix, etc. 
%we consider that fixing this part of QML platform bugs does not require ML-related knowledge.

\begin{figure*}[h]
    \centering
    \includegraphics[width=0.85\textwidth]{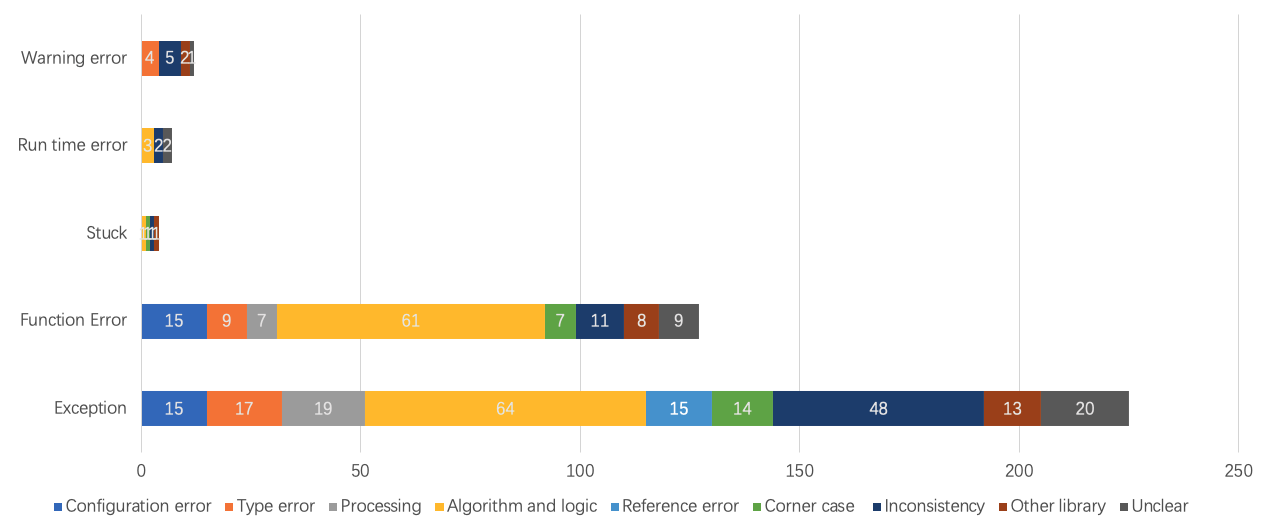}
    \caption{Distribution of root causes in each symptom.}
    \label{fig:casue}
    \vspace{-2mm}
\end{figure*}

\subsection{RQ2.Symptoms and root causes}
We classify the symptoms into the following categories:

\begin{enumerate}[itemindent=1.5em]
    \item \textbf{Function error (30.5\%).}
    A program that does not implement the appropriate function is defined as a Function error, e.g., an output error or a program that does not have an appropriate return value. 
    %An example can be seen in error \#3023\footnote{https://github.com/PennyLaneAI/pennylane/issues/3023}. 
    \item \textbf{Exception (63.2\%).}
    A program interrupts and throws exceptions, which we consider an exception.
    \item \textbf{Stuck (1.1\%).} 
    A stuck means a program keeps running without any responses.
    \item \textbf{Run time error (1.8\%).}
    This condition manifests as a program taking much longer to execute than expected. 
    \item \textbf{Warning error (3.4\%).}
    In this case, the execution of the program is usually not affected, but a warning message will be displayed, such as an outdated version, an excuse for deprecation, etc.
\end{enumerate}

\begin{figure}[h]
  \begin{CodeOut}
\footnotesize{
  \begin{alltt}
    \textcolor{purple}{-  for (int i = 0; i < programs->size(); i++)\{}
    \textcolor{teal}{+  for (size_t i = 0; i < programs->size(); i++)\{}
            Program& program = (*programs)[i];
            Status status = Status::OK();
            ...
  \end{alltt}
}
\end{CodeOut}
\Caption{\label{fig:3} Example of type error}
\end{figure}

Figure~\ref{fig:type} is an extension of \textit{Finding 1}. All the symptoms are raised within the quantum code and the traditional part. 
And the number of traditional bugs has more weight, especially for the symptom \textit{Exception}.

The causes of bugs are divided into the following categories:

\begin{enumerate}[itemindent=1.5em]
    \item \textbf{Type error (8.0\%).} 
    Bugs caused by incorrect type definitions are classified in this category. Issue \#30\footnote{https://github.com/tensorflow/quantum/issues/30} is an example caused by a type error.
    \item \textbf{Processing (7.1\%).} 
    We judge these bugs based on the quantum preparation part of the quantum circuit and the classical part related to data processing.
    \item \textbf{Inconsistency (18.6\%.)} This cause includes API and version changes which lead to bugs.
    \item \textbf{Algorithm and logic (33.8\%).} Algorithm error and Logic error are not well split. For example, many bugs result from missing functions, and we organize the problem descriptions and fix files with the idea that the functions result in illogical and incomplete algorithms.
    \item \textbf{Reference error (3.9\%).} We classify such bugs based on incorrect package import or failure to find libraries.
    \item \textbf{Corner case (6.6\%).}If a bug occurs in a particular situation, we classify it to \textit{Corner case}.
    \#1916\footnote{https://github.com/PennyLaneAI/pennylane/issues/1916} is an example.
    The bug occurs when using \CodeIn{qml.draw()} to draw a template.
    \item \textbf{Other library (6.3\%).} If a bug occurs on other libraries, we consider the bug is not relevant to the local framework. An example can be seen in issue \#557\footnote{https://github.com/PennyLaneAI/pennylane/issues/557}, the bug 
    whether worth fixing mainly depends on the restrictions on QNode~\cite{qnode} signatures.
    \item \textbf{Unclear (7.8\%).} There is no description and fixing files of one bug, nor is the discussion of the root cause as well as the fixes unclear.
    \item \textbf{Configuration error (7.9\%).} We put bugs caused by misconfigurations into this category. These bugs are mainly in the form of compilation and plug-in problems which cause the circuit to be broken or the device to be unusable, etc.
\end{enumerate}

\textbf{Finding 2.}
Regarding symptoms, the most common situation is \textit{Exception}, which means the programs cannot run smoothly. 
The second is \textit{Function error}.
Due to functional defects resulting in the output error is difficult to be found.
The lowest percentage is \textit{stuck} and \textit{Run time error}.
Besides, \textit{Warning error} would be affected by a small human factor because the environment, framework version, and compatibility between devices should be considered.

\textbf{Finding 3.}
For root causes, \textit{Algorithmic and logic} accounts for the most significant proportion, indicating that the program is not yet perfect in terms of functional implementation. There are some missing functions worth noting. Next is \textit{Inconsitency}, ranked second; in addition to bugs caused by version lag, the main thing is the updated version and rewritten features would have bugs instead. 
Other causes of bugs are more minor in proportion and evenly distributed.

\subsection{RQ3.Common challenges for developers}

\begin{figure*}[h]
    \centering
    \includegraphics[width=0.85\textwidth]{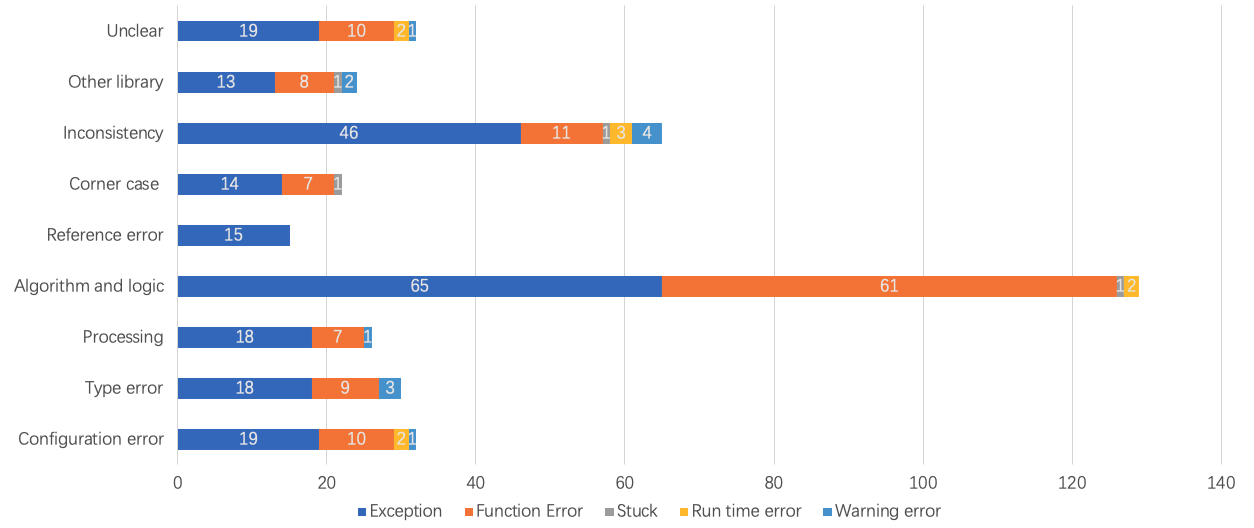}
    \caption{Distribution of symptoms in each root cause.}
    \label{fig:sym}
    \vspace{-2mm}
\end{figure*}

Figure~\ref{fig:casue} shows each symptom's distribution of root causes. The number in the figure shows the total number of bugs found.
Of all the causes we concluded, only reference errors do not result in a function error. 
And clearly, for quantum platforms, algorithms and logic errors easily cause the program to crash or get functional problems. 
Another major cause of \textit{Exception} is \textit{Inconsistency}, which indicates that API changes and version updates are most likely to cause program exceptions.
Besides, all the symptoms have \textit{Inconsistency} as a cause.

On the other hand, Figure~\ref{fig:sym} shows the distribution of symptoms in each root cause.
 \textit{Exception} has the largest share of any causal factor.
 This indicates that any of the root causes can easily lead to a program throwing exceptions.
 \textit{Function error} also accounts for a relatively large proportion,  indicating that there are also many causes of functional errors.
 \textit{Stuck}, \textit{Warning error}, and \textit{Run time error} have less weight in the root causes.% and some of them don't even appear.

\textbf{Finding 4.}
As a result, we summarized the following areas of conflict for developers:
\begin{itemize}
    \item We collected 22 repositories from nine popular organizations on GitHub. Considering the existing devices (e.g., GPUs, QVMs, QPUs, simulators) which support executing QML programs, the platform needs to provide a large number of plugins to enable cross-device execution of programs. 
    Therefore, developers need to handle the cooperation between platforms and devices to facilitate user implementation of QML programs.
    %Therefore, leveraging the cooperation between platforms and devices to implement QML programs is challenging.
    % \item \zpz{ It is common to refactor functions or methods in order to achieve functional improvements, and updated versions can have bugs that make performance inferior to previous versions. The developers could take into account the functionality of the original program while updating the version in order to prevent more bugs from appearing.}
    \item Current quantum platforms are under rapid evolution, which has a large number of undergoing and future code refactoring and compatibility changes. However, these changes are often error-prone and may cause performance degradation. Developers should pay more attention when making code changes to prevent introducing new bugs.
    %The next challenge is the version update. To achieve functional enhancements, and the phenomenon of refactoring functions or methods is more common, the updated version would appear bug, which makes the performance inferior to the previous version. 
    \item \textit{Exception} is the most common phenomenon, which means ensuring program execution is a fundamental area of conflict. In addition, the cause of many bugs can be troublesome to find as the unclear symptoms accounted for 7.8\%, sometimes requiring user assistance.
    \item Considering the \textit{Algorithm and logic} accounts for the largest share of the root cause, the consequence is many functions are incorrectly implemented. And fixing missing or defective features sometimes involves multiple source files. Therefore, fixing these bugs requires developers to have knowledge and skills related to quantum machine learning. 
\end{itemize}

\section{Threats to Validity}\label{sec:threats}

\subsection{Internal Validity}
There are some internal validities in this work.
First, we only look for bugs on GitHub, and we can't judge the audience or quality of a QML platform based only on the number of bugs we collect on GitHub. 
In our future work, more bugs would be found in other communities such as \textit{StackOverflow}, \textit{StackExchange} and \textit{Gitee}, etc.
Secondly, we did not further classify the errors from users. This is because users' misuse of QML frameworks is not this study's main topic. And most errors are caused by users not knowing how to use QML frameworks.
Finally, two authors disagreed a lot in the process of bug classification, which led to the inability to reach a high agreement rate. 

\subsection{External Validity}
Some bugs were not fixed or were too complicated to set, so we couldn't easily determine the causes. 
On the other hand, many frameworks, especially \textit{pennylane}, often combine programs with devices on other platforms. Another example, \textit{Qiskit-Machine-learning}, requires \textit{Qiskit-terra} as a base library for implementation.
This situation makes it hard to sort bugs by considering only QML libraries.

\subsection{Verifiability}
This threat is mainly about the possibility of replicating this study. We try our best to provide all related details to replicate this work. Interested researchers may find the replication package at https://github.com/Z-928/QML-platform-bugs.

\section{Related Work}\label{sec:related}
% In this section, we will discuss the following lines of closely related research.
% To the best of our knowledge, this work is the first study on bugs of quantum machine learning. 
%\subsection{Bug Benchmark Suite for Classical Software}

\subsection{Empirical Studies on Bugs in ML Systems and Frameworks}
The recent success of applying machine learning to various areas has inspired researchers to study and characterize bugs in machine learning systems extensively. Thung~\emph{et al.}~\cite{thung2012bugs} were the first to study machine learning bugs by analyzing bug reports of three machine learning systems (Apache Mahout, Lucene, and OpenNLP). Five hundred randomly picked fixed bugs were manually inspected and categorized into 11 categories. Zhang~\emph{et al.}~\cite{zhang2018issta} studied 175 TensorFlow bugs collected from GitHub and Stack Overflow (SO). They summarized the symptoms and root causes of the bugs. Humbatova~\emph{et al.}~\cite{humbatova2020icse} manually inspected bug reports from GitHub and posts in SO and proposed an extensive taxonomy of DL system faults. These studies have investigated bugs in machine learning systems, but none focuses on quantum machine learning systems. Quantum machine learning systems are unique since the quantum computation logic is expressed in quantum circuits, and the states of quantum registers are measured probabilistically. Our study will fill this research gap by performing the first systematic study on the bugs in quantum machine learning frameworks.
%Our study is to characterize the bugs in quantum machine learning frameworks systematically.

There also exist studies that look into bugs in ML frameworks. Jia~\emph{et al.}~\cite{jia2020empirical} studied the symptoms and causes of bugs in TensorFlow by analyzing 202 bugs inside TensorFlow. The authors then extend their work~\cite{jia2021symptoms} by further studying TensorFlow's repair patterns and bugs involving multiple programming languages. Chen~\emph{et al.}~\cite{chen2022bugs} conducted a large-scale study on characteristics (e.g., root cause, symptoms, and their correlations with DL framework components) of DL framework bugs by analyzing 800 bugs from four popular DL frameworks. Yang~\emph{et al.}~\cite{yang2022ist} conducted an empirical study on bug characteristics of DL frameworks by analyzing 44,083 pull requests from eight DL frameworks. However, little attention has been received to characterize bugs in QML frameworks, and our study initiates the first step in this area.

\subsection{Studies of Bugs Involve Quantum Programming}
Paltenghi and Pradel~\cite{paltenghi2022bugs} presented an empirical study of bugs in quantum computing platforms. They summarized 223 real-world bugs in some open-source quantum computing platforms, conducted a detailed statistical analysis of bugs, and proposed some bug patterns specific to the field of quantum computing platforms. 
Different from their work, this paper focuses on the bugs in QML platforms.
%\textcolor{red}{The difference between their work and ours is that they focused on the bugs in quantum computing platforms while we concentrated on the bugs in QML platforms.} 
In summary, the findings from both studies foster a deep understanding of bugs related to quantum computing systems.  

In addition, several lines of research have been carried out to study the bug patterns~\cite{zhao2021identifying,huang2019statistical}, as well as bug benchmarks~\cite{campos2021qbugs,zhao2021bugs4q} in quantum programs, to support the debugging and testing of quantum software.

\section{Conclusion}\label{sec:conclusion}
To the best of our knowledge, this work is the first study of bugs in the QML platform.
We collected 1,591 issues reported in 22 repositories, out of which we filtered out 391 bugs.
Based on the bugs found, we formulated three research questions, and four findings
%gave conclusions 
based on the analysis of the results. 
First of all, one-third of the bugs are quantum-specific bugs, which are reflected in the functionality of qubit manipulation.
Secondly, the symptom of most bugs is \textit{Crash}, followed by \textit{Function error}. These two account for 93.7\% of all bugs, while other symptoms are relatively rare. Besides, the root cause \textit{Algorithms and logic} accounts for most of the bugs, followed by \textit{Inconsistency}.
Finally, for developers, the challenges are mainly reflected in four aspects: device interaction, version change, locating symptoms, and fixing bugs. 

In future work, we would like to analyze the location of bugs and bug fixing in QML platforms, etc. In addition, there are also some bugs caused by the users of QML platforms, which are worth analyzing.

\section*{Acknowledgement} % not allowed in review
This work was supported by JST, the establishment of university fellowships towards the creation of science and technology innovation (Grant Number JPMJFS2132).

\bibliographystyle{IEEEtranS}
\bibliography{IEEEabrv, ref}

\end{document}